\documentstyle[12pt,manuscript,aps,epsf]{revtex}  
\begin{document}
\newcommand{\p}{\partial}
\newcommand{\ls}{\left(}
\newcommand{\rs}{\right)}
\newcommand{\beq}{\begin{equation}}
\newcommand{\eeq}{\end{equation}}
\newcommand{\beqa}{\begin{eqnarray}}
\newcommand{\eeqa}{\end{eqnarray}}
\newcommand{\bdm}{\begin{displaymath}}
\newcommand{\edm}{\end{displaymath}}
\newcommand{\fps}{f_{\pi}^2 }
\newcommand{\mks}{m_{{\mathrm K}}^2 }
\newcommand{\ms}{m_{{\mathrm K}}^{*} }
\newcommand{\msq}{m_{{\mathrm K}}^{*2} }
\newcommand{\rhos}{\rho_{\mathrm s} }
\newcommand{\rhob}{\rho_{\mathrm B} }
\draft
\title{Consequences of covariant kaon dynamics in heavy ion collisions} 
\author{C. Fuchs, D.S. Kosov, Amand Faessler, Z.S. Wang and T. Waindzoch}
\address{Institut f\"ur Theoretische Physik der 
Universit\"at T\"ubingen, \\
D-72076 T\"ubingen, Germany }
\maketitle
\begin{abstract}
The influence of the chiral mean field on the kaon dynamics in 
heavy ion reactions is investigated. Inside the nuclear 
medium the kaons are described as dressed quasi-particles 
carrying effective masses and momenta. A momentum dependent 
part of the interaction which resembles a 
Lorentz force originates from spatial 
components of the vector field and provides an important 
contribution to the in-medium kaon dynamics. 
This contribution is found to counterbalance the 
influence of the vector potential on the $K^+$ in-plane flow 
to a strong extent. Thus it appears to be difficult to restrict 
the in-medium potential from the analysis of the corresponding 
transverse flow.\\ \\
{\it Keywords:} Subthreshold $K^+$ production, kaon flow, 
kaon mean field, Ni+Ni, E=1.93 GeV/nucleon reaction
\end{abstract}
\pacs{25.75.+r}
In recent years strong efforts have been made towards a better understanding 
of the medium properties of kaons in dense hadronic matter. This feature 
is of particular relevance since the kaon mean field is related to 
chiral symmetry breaking \cite{kaplan86}. 
The in-medium effects give rise to an 
attractive scalar potential 
inside the nuclear medium which is in first order, i.e. in mean field 
approximation, proportional to the kaon-nucleon Sigma term 
$\Sigma_{\mathrm{KN}}$. A second part of 
the mean field originates from the interaction with vector mesons 
\cite{kaplan86,brown96,schaffner97}. The vector potential is 
repulsive for kaons $K^+$ and, due to G-parity conservation, attractive 
for antikaons $K^-$. A strong attractive potential for antikaons may 
also favor $K^-$ condensation at high nuclear densities and thus modifies 
the properties of neutron stars \cite{li97b}. 

One has extensively searched for signatures of 
these kaon-nucleus potentials, in particular in heavy ion reactions at 
intermediate energies \cite{kaos94,fopi95,kaos97,fopi97}. 
Here the transverse flow of $K^+$ mesons has 
attracted special attention. Due to the lack of reabsorption, $K^+$ 
mesons should, in contrast to $K^-$ mesons and pions, 
not exhibit a pronounced 
shadowing effect but rather reflect the behavior of the primary 
sources, i.e. the nucleons (or $\Delta$--resonances). However, as first 
proposed by Li and Ko \cite{ko95} the presence of a $K^+$ potential 
should distort this scaling of the flow. The repulsive vector 
potential tends to push the $K^+$ mesons 
away from the nucleons leading to an anti-correlated 
flow. The attractive scalar potential counterbalances this 
behavior to some extent. Thus the net effect is expected to 
be a zero flow around midrapidity. 

Due to its relativistic origin, the kaon mean field has a typical 
relativistic scalar--vector type structure. For the nucleons such 
a structure is well known from Quantum Hadron Dynamics \cite{serot88}.  
This decomposition of the mean field 
is most naturally expressed by an absorption of the scalar and vector 
parts into effective masses and momenta, respectively, leading to 
a formalism of quasi-free particles inside the nuclear medium
\cite{serot88}. The application of the quasi-particle picture in 
heavy ion collisions is well established for nucleons and has 
extensively been used within 
the framework of relativistic transport theories 
\cite{fuchs96b,ko97a,cassing97a}. 

In the present work we extend 
this manifestly covariant approach to the treatment of 
kaon in-medium properties and discuss 
implications on the $K^+$ flow in heavy ion reactions. 
Starting from the chiral Lagrangian set up by Kaplan and Nelson 
\cite{kaplan86} Li and Ko investigated the kaon properties in mean 
field approximation \cite{ko95}. Including higher order terms 
kaon potentials have been derived within chiral perturbation theory 
by Brown and Rho \cite{brown96} 
and by Waas, Weise and Kaiser \cite{waas96}. 
Our starting point will be the mean field approximation 
\cite{ko95}. 

From the chiral Lagrangian the field equations for 
the $K^\pm$--mesons are derived from the 
Euler-Lagrange equations \cite{ko95} 
\beq
\left[ \partial_\mu \partial^\mu \pm \frac{3i}{4\fps} j_\mu \partial^\mu 
+ \left( \mks - \frac{\Sigma_{\mathrm{KN}}}{\fps} \rhos \right) 
\right] \phi_{\mathrm{K^\pm}} (x) = 0
\quad .
\label{kg1}
\eeq
Here the mean field approximation has already been applied. In Eq. 
(\ref{kg1}) $ \rhos$ is the baryon scalar density, $j_\mu$ the baryon 
four-vector current and $i=\sqrt{-1}$ ensures the hermiticity of the 
operator. To make Eq. (\ref{kg1}) more transparent 
the kaonic vector potential 
\beq 
V_\mu = \frac{3}{8\fps} j_\mu
\label{vpot1}
\eeq
is introduced and Eq. (\ref{kg1}) is rewritten in the form 
\beqa
&&\left[ \left( \partial_\mu + i V_\mu \right)^2  + \msq \right] 
\phi_{\mathrm{K^+}} (x) = 0 
\label{kg2a}\\
&&\left[ \left( \partial_\mu - i V_\mu \right)^2  + \msq \right] 
\phi_{\mathrm{K^-}} (x) = 0 
\quad . 
\label{kg2b}
\eeqa
Thus, the vector field is introduced by minimal coupling 
into the Klein-Gordon equation. The effective mass $\ms$ of 
the kaon is then given by 
\beq
\ms = \sqrt{ \mks - \frac{\Sigma_{\mathrm{KN}}}{\fps} \rhos 
     + V_\mu V^\mu }
\quad . 
\label{mstar1}
\eeq
Due to the bosonic character, the coupling of the 
scalar field to the mass term is no longer linear 
as for the baryons but quadratic and contains an additional contribution 
originating from the vector field. The effective quasi-particle 
mass defined by Eq. (\ref{mstar1}) is a Lorentz scalar and 
is equal for $K^+$ and $K^-$. It should not be mixed up with 
the quantity, i.e. kaon energy at zero momentum, which is 
sometimes denoted as in-medium mass 
\cite{ko97a,cassing97a,waas96,cassing97b} and which determines the 
shift of the corresponding production thresholds. The form of Eqs. 
(\ref{vpot1}-\ref{mstar1}) is quite general and also holds for the 
description of other spin-0 mesons, i.e. pions, where 
the mesons interact with an external vector field, respectively 
the baryon field, by a derivative coupling in the 
Lagrangian. Introducing an effective momentum 
\beq
k^{*}_{\mu} = k_{\mu} \mp V_\mu
\label{keff}
\eeq
for $K^+ (K^- )$, the Klein-Gordon 
equation (\ref{kg2a},\ref{kg2b}) reads in momentum space 
\beq
\left[ k^{*2} - \msq \right] \phi_{\mathrm{K}^\pm } (k) = 0
\label{kg3}
\eeq
which is just the mass-shell constraint for the quasi-particles 
inside the nuclear medium. These quasi-particles can now 
be treated like free particles. Such a description is 
analogous to the well known quasi-particle picture for 
nucleons in a relativistic mean field \cite{serot88} where 
the in-medium spinors obey an effective Dirac equation 
$ \left[ {\not k}^* - m^* \right] u^* (k) = 0$. 
In nuclear matter at rest the spatial components of the 
vector potential vanish, i.e. ${\bf V} = 0$, and 
Eqs. (\ref{kg2a},\ref{kg2b}) reduce to the expression already 
given in Ref. \cite{ko95}. 
However, Eqs. (\ref{kg2a},\ref{kg2b}) generally account for the 
correct Lorentz properties which are not obvious from the 
standard treatment of the kaon 
mean field \cite{ko95,ko97a,cassing97a,li97b}. 
In particular the quadratic term in density which enters into 
the effective mass is a Lorentz scalar determined by the baryon 
density in the local nuclear matter 
rest frame (RF), i.e. the frame where the spatial components of 
the brayon current vanishes,
\beqa
V_\mu V^\mu = \left(\frac{3}{8\fps}\right)^2 j_\mu j^\mu 
= \left(\frac{3}{8\fps}\right)^2
\left( \rhob |_{\mathrm{RF}} \right)^2 
\quad .
\label{vsq}
\eeqa

To keep the present investigations as general as possible various 
mean field parameterizations with different strengths are 
considered. With MF we denote the mean field obtained with 
a value of $\Sigma_{\mathrm{KN}}$=350 MeV 
originally used in \cite{ko95}. For comparison we 
also adopt a mean field suggested by Brown and Rho \cite{brown96} 
which has a value of $\Sigma_{\mathrm{KN}}$=450 MeV and accounts 
for the rescaling of the 
vector part by medium modifications to the pion decay constant, 
i.e. $f_{\pi}^{*2} = 0.6 f_{\pi}^{2}$. The second interaction, 
denoted as MF2 in the following, results in a much stronger 
vector part which is nearly twice as large as for MF. 
The MF2 parameterization has already been used in our 
previous investigations based on 
the non-covariant QMD approach \cite{wang97,wang97b}.

Higher order corrections to the kaon self-energy have been 
found, e.g., in Ref. \cite{waas96} 
to cancel the attractive part of the scalar 
self-energy given by $ -\Sigma_{\mathrm{KN}} / \fps \rhos$ to high 
extent. In order to complete the systematics 
the results of Ref. \cite{waas96} 
are parameterized on the mean field level. 
Motivated by Eqs. (\ref{kg2a},\ref{kg2b}) we make the 
following ansatz for the dispersion relation given in the nuclear matter 
rest frame
\beqa
\omega_{\mathrm{K^\pm}} ({\bf k}=0) = \ms \pm V_0 
\quad , \quad 
\ms = \sqrt{ \mks - {\tilde \Sigma}_{\mathrm{S}}^2  + V_{0}^2 } 
\quad .
\label{para}
\eeqa
The quantities 
$ V_0 = \frac{1}{2} \left( \omega_{\mathrm{K^+}}
- \omega_{\mathrm{K^-}}\right)$ and 
$\ms = \frac{1}{2} \left( \omega_{\mathrm{K^-}}
+ \omega_{\mathrm{K^+}}\right)$ are determined from the 
dispersion relations. To account for the correct density 
dependence scalar and vector parts are represented by non-linear 
proportionality factors $V_\mu = g_{{\rm V}} (\rhob) j_\mu$ 
and ${\tilde \Sigma}_{\mathrm{S}}^2 = g_{{\rm S}} (\rhob) \rhos$ 
which defines the coupling functions $g_{{\rm V}} =V_0 /\rhob$ and 
$ g_{{\rm S}} = {\tilde \Sigma}_{\mathrm{S}}^2 /\rhos$. 
This parameterization reproduces the results 
of Ref. \cite{waas96} with high accuracy. 

In Fig. 1 the effective quasi-particle masses $\ms$ obtained in the 
mean field (MF, MF2) approach, Eq. (\ref{mstar1}), and from chiral 
perturbation theory (ChPT) Ref. \cite{waas96},  
Eq. (\ref{para}), are compared. 
We want to stress again that the 
quasi-particle mass $\ms$ is equal for $K^+$ and $K^-$. The effective 
mass $\ms$ is 
generally reduced inside the nuclear medium due to the attractive 
scalar potential. However, in the approach of Ref. \cite{waas96} 
(ChPT) this reduction is much weaker than in the simple MF 
parameterization where the scalar field is in first order proportional 
to the scalar nucleon density. Interestingly enough, the effective mass 
in the MF2 parameterization behaves similar to MF which is due 
to the fact that the significantly higher value of 
$ \Sigma_{\mathrm{KN}}$ is counterbalanced 
by the stronger quadratic vector term.  

The covariant equations of motion are obtained in the classical 
(testparticle) limit from the relativistic transport 
equation for the kaons which can be derived from Eqs. 
(\ref{kg2a},\ref{kg2b}). They are analogous to the 
corresponding relativistic equations for baryons and read  
\beqa
\frac{ d  q^\mu}{d\tau} = \frac{k^{*\mu}}{\ms}
\quad , \quad 
\frac{ d  k^{*\mu}}{d\tau} = \frac{k^{*}_{\nu}}{\ms} F^{\mu\nu} 
+\partial^\mu \ms
\quad . 
\label{como}
\eeqa
Here $q^\mu = (t,{\bf q})$ are the coordinates in Minkowski space 
and $F^{\mu\nu} = \partial^\mu  V^\nu - \partial^\nu  V^\mu $ is the 
field strength tensor for $K^+$. For $K^-$ where the vector field 
changes sign the equation of motion are identical, however, 
$F^{\mu\nu}$ has to be replaced by $-F^{\mu\nu}$. 
The structure of Eqs. (\ref{como}) may become 
more transparent considering only the spatial components
\beq
\frac{d {\bf k^*}}{d t} = - \frac{\ms}{E^*} 
\frac{\partial \ms }{\partial {\bf q}} \mp 
\frac{\partial V^0 }{\partial {\bf q}} 
\pm \frac{{\bf k}^*}{E^*} \times 
\left( \frac{\partial}{\partial {\bf q}} \times {\bf V} \right)
\label{lorentz}
\eeq
where the upper (lower) signs refer to $K^+$ ( $K^-$). 
The term proportional to the spatial component of the vector 
potential gives rise to a momentum dependence 
which can be attributed to a Lorentz force, i.e. 
the last term in Eq. (\ref{lorentz}). Such a velocity dependent 
$({\bf v} = {\bf k}^* / E^* )$ Lorentz force 
is a genuine feature of relativistic dynamics as soon 
as a vector field is involved. Concerning the nucleons such 
terms are generally included in relativistic transport approaches 
\cite{fuchs96b,cassing97a,cassing97b} and have been found to 
provide an essential contribution to the dynamics 
\cite{fuchs96b,blaettel93}. 

If the equations of motion are, however, derived from a static 
potential 
\beqa
 U({\bf k},\rho) = \omega({\bf k},\rho) - \omega_0 ({\bf k}) 
= \sqrt{{\bf k}^2 +  \mks - \frac{\Sigma_{\mathrm{KN}}}{\fps} \rhos 
     + V_{0}^2 } \pm V_0 - \sqrt{{\bf k}^2 +  \mks }
\label{pot}
\eeqa
as, e.g. in Refs. \cite{li97b,ko95,ko97a,cassing97b}, the 
Lorentz-force like contribution is missing. The same holds for 
non-relativistic approaches \cite{wang97,wang97b} where 
the Lorentz force has also not yet been taken into account. 
Such non-covariant treatments are formulated 
in terms of canonical momenta $k$ instead of kinetic 
momenta $k^*$ and then the equations of motion (\ref{lorentz}) 
read
\begin{equation}
\frac{d {\bf k}}{d t} = - \frac{\ms}{E^*} 
\frac{\partial \ms }{\partial {\bf q}} \mp 
\frac{\partial V^0 }{\partial {\bf q}} 
\pm {\bf v}_{i} \frac{\partial  {\bf V}_{i}}{\partial {\bf q}} 
\quad ,
\label{lorentz2}
\end{equation}
with ${\bf v} = {\bf k}^* / E^*$ the kaon velocity. 

In the following the influence of covariant dynamics is 
examined for $K^+$. Since in particular the in-plane flow 
turned out to react sensitively on the kaon mean field 
\cite{ko95,ko97a,wang97} we will focus 
on this observable. We consider the reaction 
Ni on Ni at 1.93 A.GeV incident energy which has been studied 
extensively from the theoretical \cite{ko97a,cassing97b,wang97} 
as well as from the experimental \cite{fopi95,fopi97} side. 
The $K^+$ creation mechanism is thereby 
treated as described in Ref. \cite{fuchs97b}. However, in contrast to 
Ref. \cite{fuchs97b}, we use the improved cross section of 
Ref. \cite{sibirtsev95} for the baryon induced $K^+$ creation channels; 
for the pion induced channels we adopt the cross sections 
given in Refs. \cite{tuebingen1}. 
The baryon dynamics are treated within the framework 
of Relativistic Quantum Molecular Dynamics (RQMD). Here the 
extension of this approach to covariant 
scalar--vector mean fields \cite{fuchs96b} is used. For the 
nucleon mean field we adopt the 
non-linear NL2 version of the $\sigma\omega$ model which is able to 
reasonably describe the dynamical features of heavy ion collisions 
in the energy range considered \cite{fuchs96b,ko97a}. 

In Fig. 2 exemplarily a semi-central (b=3 fm) reaction is 
considered. For better transparency of the results the 
rescattering of $K^+$ mesons with baryons is not included 
in these calculations since it may wash out the mean field effects 
to some extent \cite{nantes97}. The standard treatment 
(upper panel) where kaons are propagated in the 
static potential, Eq. (\ref{pot}), 
is compared to the covariant approach (lower panel) 
given by Eq. (\ref{como}), respectively Eq. (\ref{lorentz}). 
Without any potential a clear flow signal is observed which reflects 
the transverse flow of the primary sources of the $K^+$ production. 
The dominantly repulsive character of the in-medium potential 
(\ref{pot}) tends to push the kaons away from the spectator matter 
which leads to a zero flow around midrapidity for the ChPT and 
also a nearly vanishing flow for the MF mean field. 
This deviation can be understood from the fact that the 
magnitude of the vector field is stronger using the 
ChPT parameterization ($V_0 = 79$ MeV 
at a nuclear density of 0.16 $fm^{-3}$) compared to MF 
($V_0 = 57$ MeV). The effective mass, Fig. 1, starts, however, 
to differ significantly only at higher densities. Using the more 
repulsive MF2 parameterization the vector potential is so 
strong ($V_0 = 95$ MeV) that it produces an anti-correlated 
flow signal which is in good agreement with our previous 
investigation using the QMD model \cite{wang97} and 
also with the results of Ref. \cite{ko97a} 

The situation changes, however, 
dramatically when the full Lorentz structure of the mean field 
is taken into account. The influence of the repulsive part of the 
potential, i.e. the time-like component, on the in-plane flow is 
almost completely counterbalanced by the velocity dependent part 
of the interaction. Hence, no net effect of the potential is 
any more visuable. This feature is rather independent 
on the actual strength of the potential, i.e. it is the same 
for all potentials considered. Although the Lorentz force vanishes in nuclear 
matter at rest, it is clear that this force generally 
contributes in heavy ion collisions. 
Kaons are produced in the early phase of the reaction where the 
relative velocity of projectile and target matter is large. Thus 
the kaons feel a non-vanishing baryon current in the spectator 
region, in particular in non-central collisions.

In Fig. 3 the calculations are finally compared to the 
FOPI data \cite{fopi95}. The results are obtained for impact 
parameters $b\leq 4$ fm and with a transverse momentum cut 
$P_{{\rm T}} / m_{{\rm K}} > 0.5$. Here the rescattering 
of the $K^+$ mesons with the baryons and the influence of 
the Coulomb force is taken into account. The electromagnetic 
interaction is treated analogously to the strong interaction, i.e. 
by adding $F^{\mu\nu}_{{\rm el }} 
= \partial^\mu  A^\nu - \partial^\nu  A^\mu $ given by the 
electromagnetic vector potential to Eq. (\ref{como}). It again turns 
out that the influence of the covariant in-medium dynamics 
on the $K^+$ in-plane flow is hardly visuable. The calculation 
including the mean field (MF) yields a $K^+$ flow which 
is very close to the result obtained without 
any potential. The latter is in good agreement with the 
calculation of Ref. \cite{ko97a} for the free case. However, 
comparing to the experiment it seems that in the rapidity range 
$| Y_{{\rm c.m.}} / Y_{{\rm proj}}| \sim 0.5 - 1$ 
the flow is slightly overpredicted with respect to the data 
although the calculations lie inside the error limits. 
The more sophisticated ChPT potential leads to a slight 
reduction of the flow and thus to an 
improved agreement with the data which is, however, 
not satisfying. This is in particular the case if one has 
the preliminary and yet unpublished new flow data with smaller error 
bars in mind. 

The cancelation effects on the flow can be 
understood from Eq. (\ref{lorentz2}). 
The vector field is generally 
proportional to the baryon current 
$ j_\mu = (\rho_B, {\bf u}\rho_B) $ where ${\bf u}$ denotes the 
streaming velocity of the surrounding nucleons. Let us for the 
moment assume that ${\bf u}$ is locally constant, then the 
total contribution of the vector field in Eq. (\ref{lorentz2}) 
can be written as 
$ \mp \frac{3}{8\fps} 
\left( 1 - |{\bf v}| |{\bf u}| \cos\Theta \right) 
\frac{\partial \rho_B }{\partial {\bf q}} 
$. 
Now the angle $\Theta$ between the kaon and the baryon 
streaming velocities determines the influence of the 
Lorentz force. Since in our case and also in the 
calculations of Refs. \cite{ko95,ko97a} the $K^+$s initially 
follow the primordial flow of the nucleons we have 
$\cos\Theta \sim 1$ which gives rise to the cancelation. 
However, the value of $\Theta$ and also the magnitude of the 
kaon velocity are also related 
to the rescattering of the  $K^+$ mesons with the nucleons. 
An enhanced rescattering as well as a different primordial 
$K^+$ flow might reduce the cancelation 
effects from the Lorentz force. Hence, the complete description 
of the in-plane $K^+$ flow is still an open question and further 
theoretical studies seem to be necessary.  
 
In the present work the influence of the chiral mean field 
on the kaon dynamics in heavy ion reactions has been investigated. 
Three types of in-medium potentials with different strength, 
i.e. simple mean field parameterizations and a more sophisticated 
potential derived from 
chiral perturbation theory have been applied. The kaons are 
thereby described as quasi-particles carrying effective masses and 
momenta. This accounts for the correct mass-shell properties of the 
particles inside the nuclear medium. Due to the bosonic character of the 
kaons the corresponding mass-shell condition is slightly different 
to that of the nucleons. The most striking consequence 
of covariant kaon dynamics is, however, the appearance of a momentum 
dependent force proportional to the spatial components 
of the vector field which resembles the Lorentz force 
in electrodynamics. Although such Lorentz forces vanish in 
equilibrated nuclear matter they provide an essential contribution 
to the dynamics in the case of energetic heavy ion collisions. 
To summarize, the influence of in-medium effects on the in-plane 
flow is counterbalanced to large extent by this additional 
contribution and the improved agreement with the present flow 
data is destroyed. Concerning the flow of $K^-$ mesons the 
influence of the momentum dependent Lorentz force which in 
principle also counteracts against the static 
time-like component might be changed due to the dominance of 
reabsorption processes. The restriction of the kaon in-medium properties 
by the measurement of the in-plane flow appears to be 
difficult, as well for $K^+$ as for $K^-$. A way out of this problem 
may be the study of radial flow phenomena \cite{wang98}. The radial flow 
can be expected to be less distorted by the Lorentz-force terms 
since it mainly occurs in the fireball region. However, the 
present results indicate that a realistic 
description of the kaon dynamics requires to go beyond the 
-- only density dependent -- mean field approximation. E.g., 
the explicit momentum dependence of the kaon-nucleus 
potential which arises from p-wave contributions in the 
KN interaction should be included in future investigations.  


\begin{figure}
\begin{center}
\leavevmode
\epsfxsize = 15cm
\epsffile[0 50 400 350 ]{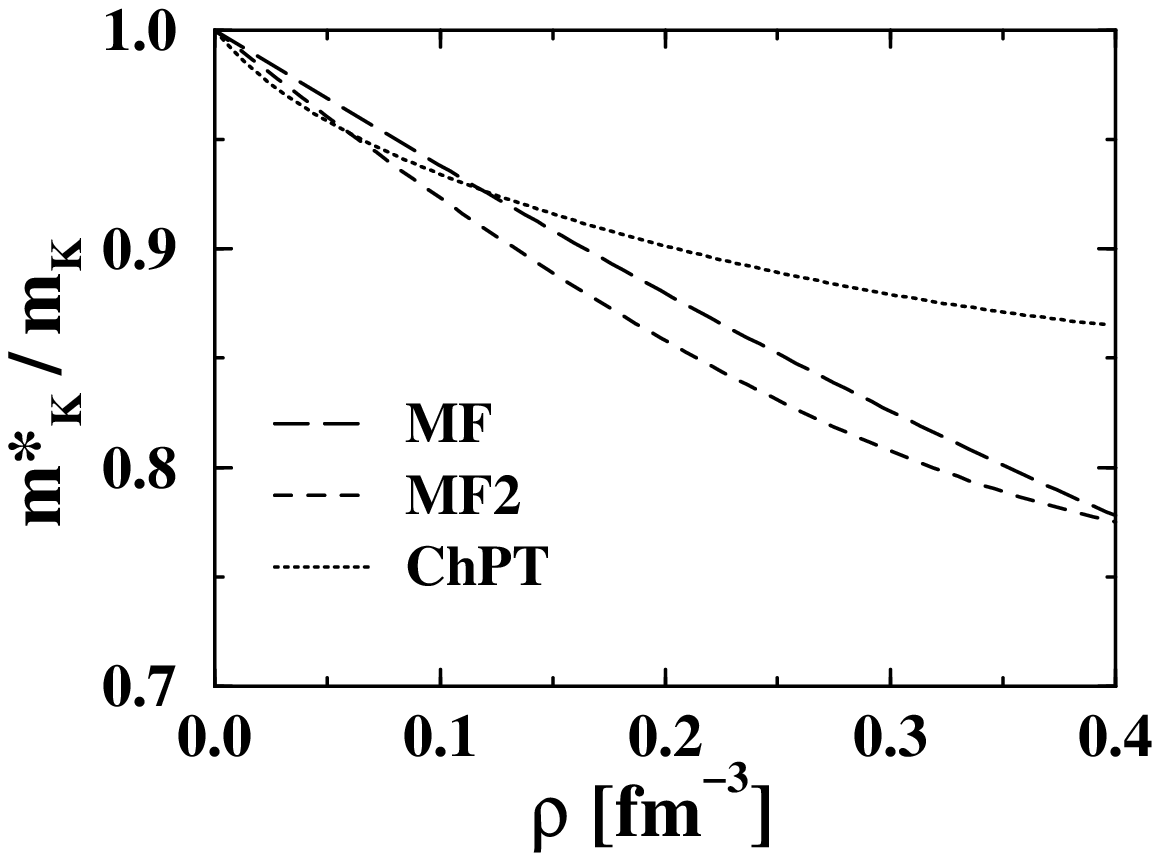}
\end{center}
\caption{Effective kaon (quasi-particle) mass in nuclear matter. 
The mean field parameterizations MF (long-dashed) and 
MF2 (dashed) are compared to the result from 
chiral perturbation theory \protect\cite{waas96} ChPT (dotted).
}
\label{mstarfig}
\end{figure}
\begin{figure}
\begin{center}
\leavevmode
\epsfxsize = 15cm
\epsffile[30 100 400 500 ]{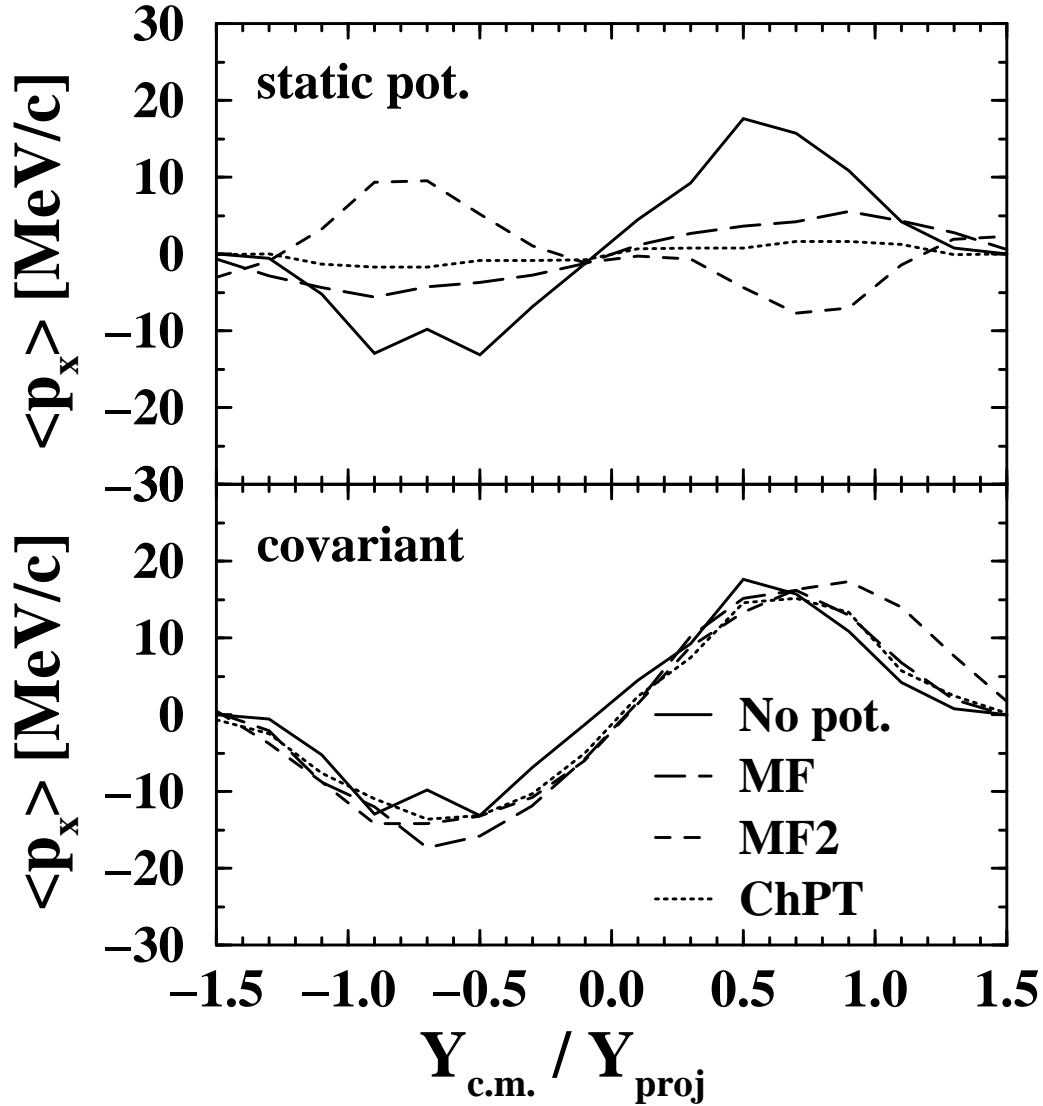}
\end{center}
\caption{Influence of a covariant description on the 
transverse $K^+$ flow in a semi-central (b=3 fm) Ni + Ni 
collision at 1.93 A.GeV incident energy. 
The calculations are performed without (solid lines) and including 
different $K^+$ mean fields MF (long-dashed) and MF2 (dashed) 
as well as the respective one 
derived from chiral perturbation theory (ChPT, dotted). 
The full covariant dynamics including Lorentz forces (bottom) 
are compared to the description with a static potential (top). 
}
\label{mstarfig}
\end{figure}
\begin{figure}
\begin{center}
\leavevmode
\epsfxsize = 15cm
\epsffile[30 100 400 460 ]{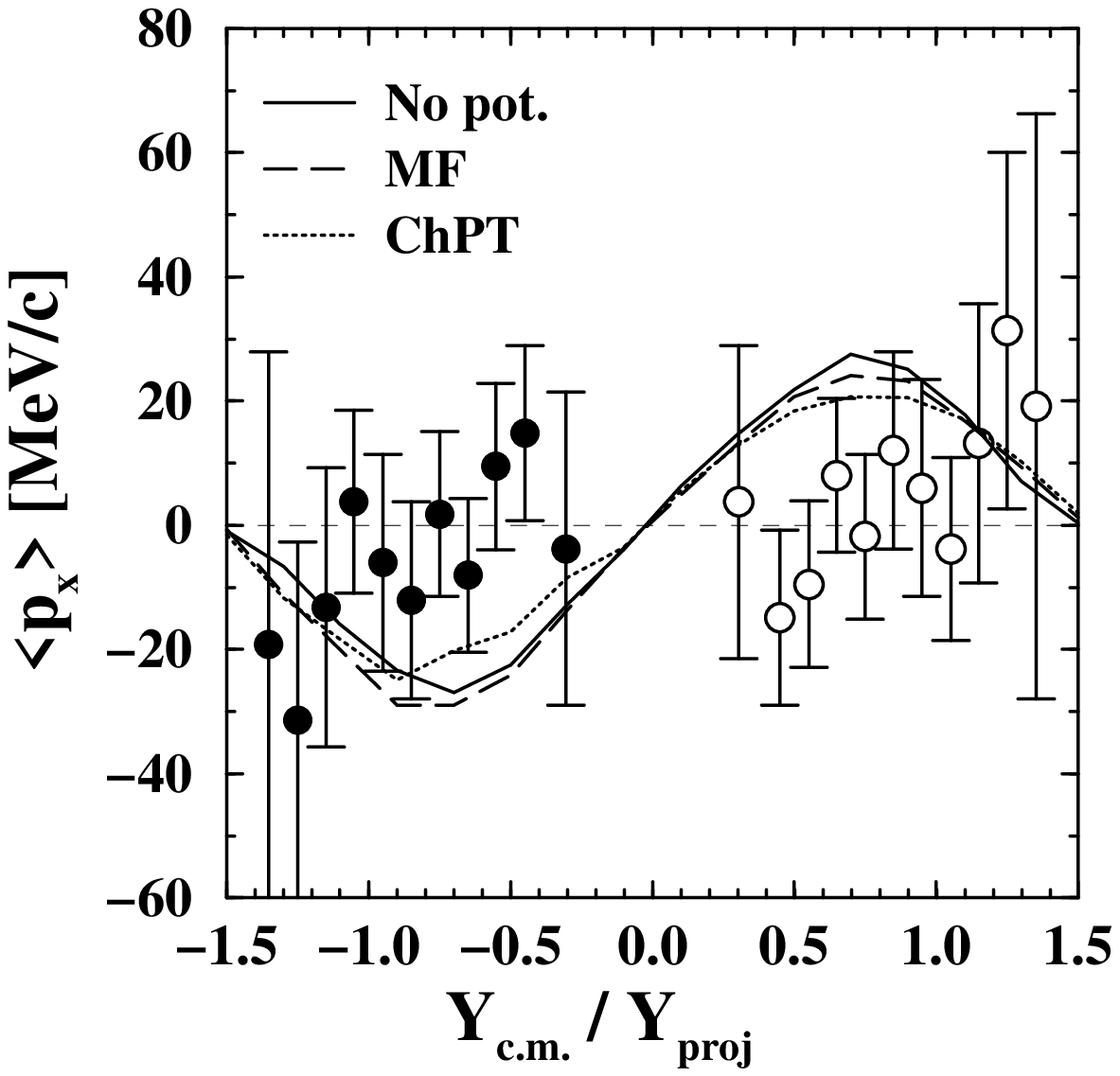}
\end{center}
\caption{Average $K^+$ in-plane flow in a Ni + Ni collision 
(b$\leq 4$ fm) at 1.93 A.GeV incident energy. 
The calculations are performed without (solid lines) and including 
a $K^+$ mean field (MF, long-dashed) and the respective one 
derived from chiral perturbation theory (ChPT, dotted). 
A transverse momentum cut $P_{{\rm T}} / m_{{\rm K}} > 0.5$ has 
been applied. The FOPI data \protect\cite{fopi95} are shown by 
circles.  
}
\label{mstarfig}
\end{figure}

\begin{thebibliography}{99} 

\bibitem{kaplan86}
B. D. Kaplan and A. E. Nelson, Phys. Lett. B 175 (1986) 57;\\
A. E. Nelson and B. D. Kaplan, Phys. Lett. B 192 (1987) 193.

\bibitem{brown96}
G.E. Brown and M. Rho, Nucl. Phys. A 596 (1996) 503.

\bibitem{schaffner97}
J. Schaffner, J. Bondorf, I. N. Mishustin, 
Nucl. Phys. A 625 (1997) 325.

\bibitem{li97b}
G.Q. Li, C.-H. Lee, G.E. Brown, Nucl. Phys. A 625 (1997) 372;\\
Phys. Rev. Lett. 79 (1997) 5214.

\bibitem{kaos94}
D. Miskowiec and the KaoS Collaboration, Phys. Rev. Lett. 72 
(1994) 3650. 

\bibitem{fopi95}
J.L. Ritman and the FOPI Collaboration, 
Z. Phys. A 352 (1995) 355;\\
D. Best, in {\em Advances in Nuclear Dynamics}, ed. 
by W. Bauer et al., (World Scientific, Singapore, 1996).

\bibitem{kaos97}
R. Barth and the KaoS Collaboration, Phys. Rev. Lett. 78 (1997) 4007. 

\bibitem{fopi97}
D. Best and the FOPI Collaboration, Nucl. Phys. A 625 (1997) 307.

\bibitem{ko95}
G. Q. Li, C. M. Ko, and Bao-An Li, Phys. Rev. Lett. 74 (1995) 235;\\
G. Q. Li, C. M. Ko, Nucl. Phys. A 594 (1995) 460.

\bibitem{serot88} 
B.D. Serot and J.D. Walecka, Adv. Nucl. Phys. 16 (1988) 1.

\bibitem{fuchs96b}
C. Fuchs, E. Lehmann, L. Sehn, F. Scholz, J. Zipprich, T. Kubo, and 
Amand Faessler, Nucl. Phys. A 603 (1996) 471.

\bibitem{ko97a}
C.M. Ko, G.Q. Li, J. Phys. G 22 (1996) 1673.

\bibitem{cassing97a}
W. Cassing, E. L. Bratkovskaya, U. Mosel, S. Teis, A. K. Weber, 
A. Sibirtsev, Nucl. Phys. A 614 (1997) 415.

\bibitem{waas96}
T. Waas, N. Kaiser, and W. Weise, Phys. Lett. B 379 (1996) 34.

\bibitem{cassing97b}
E. L. Bratkovskaya, W. Cassing and U. Mosel, 
Nucl. Phys. A 622 (1997) 593.

\bibitem{blaettel93}
B. Bl\"attel, V. Koch and U. Mosel, 
Rep. Prog. Phys. 56 (1993) 1.

\bibitem{wang97}
 Z. Wang, Amand Faessler, C. Fuchs, V.S. Uma 
Maheswari, D.S. Kosov, Nucl. Phys. A 628 (1997) 151. 

\bibitem{wang97b}
 Z. Wang, Amand Faessler, C. Fuchs, V.S. Uma 
Maheswari, D.S. Kosov, Phys. Rev. Lett. 79 (1997) 4096.

\bibitem{fuchs97b}
C. Fuchs, Z. Wang, L. Sehn, Amand Faessler, V.S. Uma 
Maheswari, D.S. Kosov, Phys. Rev. C 56 (1997) R606. 

\bibitem{sibirtsev95}
A. Sibirtsev, Phys. Lett. B 359 (1995) 29. 

\bibitem{tuebingen1}
K. Tsushima, S. W. Huang, A. Faessler, Phys. Lett. B 337 (1994) 245;\\
J. Phys. G 21 (1995) 33.

\bibitem{nantes97} 
C. David, C. Hartnack, M. Kerveno, J.-CH. Le Pallec and 
J. Aichelin, preprint nucl-th/9611016.

\bibitem{wang98}
 Z.S. Wang, Amand Faessler, C. Fuchs, V.S. Uma 
Maheswari, and T. Waindzoch, Phys. Rev. C 57 (1998) 3284.


\end{thebibliography}
\end{document}